\documentclass[aps,pre,twocolumn]{revtex4-1}
\usepackage{amssymb}
\usepackage{amsmath,bm}
\usepackage{graphicx,color}
\usepackage{bbm}
\usepackage[bottom]{footmisc}
\usepackage{multirow}
\newcommand{\B}[1]{{\bm{#1}}}

\begin{document}
\title{Fatigue and Collapse of Cyclically Bent Strip of Amorphous Solid}
\author{Bhanu Prasad Bhowmik$^1$, 
	H.G.E.Hentschel$^{1,2}$ and Itamar Procaccia$^{1,3}$}
\affiliation{$^1$Dept. of Chemical Physics, The Weizmann Institute of
	Science, Rehovot 76100, Israel\\
$^2$ Dept. of Physics, Emory University, Atlanta Ga. 30322, USA\\$^3$  Center for OPTical IMagery Analysis and Learning, Northwestern Polytechnical University, Xi'an, 710072 China. }

\begin{abstract}
Fatigue caused by cyclic bending of a piece of material, resulting in its mechanical failure, is a phenomenon that had been studied for
ages by engineers and physicists alike. In this Letter we study such fatigue in a strip of athermal amorphous solid. On the basis of
atomistic simulations we conclude that the crucial quantity to focus on is the {\em accumulated damage}. Although this quantity exhibits
large sample-to-sample fluctuations, its dependence on the loading determines the statistics of 
the number of cycles to failure. Thus we can provide a scaling theory for the W\"ohler plots of mean number of cycles for failure as a function of the
loading amplitude. 
\end{abstract}
\maketitle

{\bf Introduction:} It is known
to every child that bending back and forth a metal wire results in a final breaking of the wire. Engineers were concerned with the
fatigue that results from oscillatory and randomly applied strains for many decades \cite{54Mil,66Dow,89DBS,94KC}. Very well known are the so-called ``W\"ohler diagrams" or ``s-n plots", which display the exponential decrease of the number of cycles for failure upon increasing the maximal external load.  This exponential dependence is known since the days of W\"ohler who investigated the famous 1842 train crash in Versailles, France, but the origin of this dependence remained obscure until now. In spite of decades of research, riddles abound. In an early paper Freudenthal, Gumbel and Gough \cite{53FGG}
wrote as follows: ``The interpretation of the results of engineering tests of specimens subjected to repeated load cycles is made difficult by the fact that progressive damage, which finally leads to fatigue failure, is a process that is essentially determined by happenings on the submicroscopic scale; on a phenomenological scale its cumulative effect becomes visible only at such an advanced stage of the test at which most of the damage  has  already  been  done.  Because  of the  structure-sensitivity  of such a process, the results of fatigue tests show a much wider scatter than the results of any other mechanical test".
And continuing: ``Practically all existing fatigue theories ... operate on the assumption that fatigue can be explained in terms of a single mechanism. The fact is not considered that one mechanism alone can hardly be expected to describe a phenomenon that is the result of force- and time-dependent processes on the microscopic and submicroscopic level, which are associated with the existence of highly localized textural stress fields, defects and anomalies in the ideal structure of the material.The usual engineering abstraction of such a material as a continuous, homogeneous, isotropic, elastic body therefore precludes any effective theoretical approach to fatigue." In this Letter we indeed follow up on these comments. We employ atomistic simulations of a strip of athermal amorphous solid to precisely focus on ``progressive damage", on its statistics and on its dependence on the amplitude of the load. This allows us to provide an understanding of the dependence of the mean number of cycles to failure, as a function of the load amplitude (W\"ohler diagrams). Rather than focusing on detailed mechanistic studies, which appear to lead nowhere, we will use scaling concepts to provide a predictive theory based on a minimal number of material properties. 

{\bf Simulations:} Our model system is made of ternary mixture of point particles A, B and C with a concentration ratio A:B:C =  54:29:17, which is embedded in two dimensions. We choose a ternary system to allow a deep quench using Swap Monte Carlo, and see \cite{20POB} for details. The particles interact via a Lennard-Jones potential-
\begin{equation}
	V_{\alpha,\beta} = 4\epsilon_{\alpha\beta} [(\sigma_{\alpha\beta}/r)^{12} - (\sigma_{\alpha\beta}/r)^6], 
\end{equation}
where $\alpha$ and $\beta$ stand for different types of particle. The potential vanishes at $r_c = 1.75\sigma_{\alpha,\beta}$. This value of $r_c$ is chosen to make the material neither too ductile nor too brittle\cite{11DKPZ}. The energy scales are $\epsilon_{AB}$ = $1.5\epsilon_{AA}$, $\epsilon_{BB}$ = $0.5\epsilon_{AA}$, $\epsilon_{AC}$ = $0.5(\epsilon_{AA} + \epsilon_{AB})$, $\epsilon_{BC}$ = $0.5(\epsilon_{AB} + \epsilon_{BB})$ and $\epsilon_{CC}$ = $0.5(\epsilon_{AA} + \epsilon_{BB})$, with $\epsilon_{AA}$ equal 1.
The ranges of interaction are $\sigma_{AB}$ = $0.8\sigma_{AA}$, $\sigma_{BB}$ = $0.88\sigma_{AA}$, $\sigma_{AC}$ = $0.5(\sigma_{AA} + \sigma_{AB})$, $\sigma_{BC}$ = $0.5(\sigma_{AB} + \sigma_{BB})$ and $\sigma_{CC}$ = $0.5(\sigma_{AA} + \sigma_{BB})$, with $\sigma_{AA}$=1. The mass $m$ of each particle is unity, and the unit of time is $\sqrt{m\sigma_{AA}^2/\epsilon_{AA}}$. Boltzmann's constant is taken as unity. 

\begin{figure}
	\includegraphics[scale=0.45]{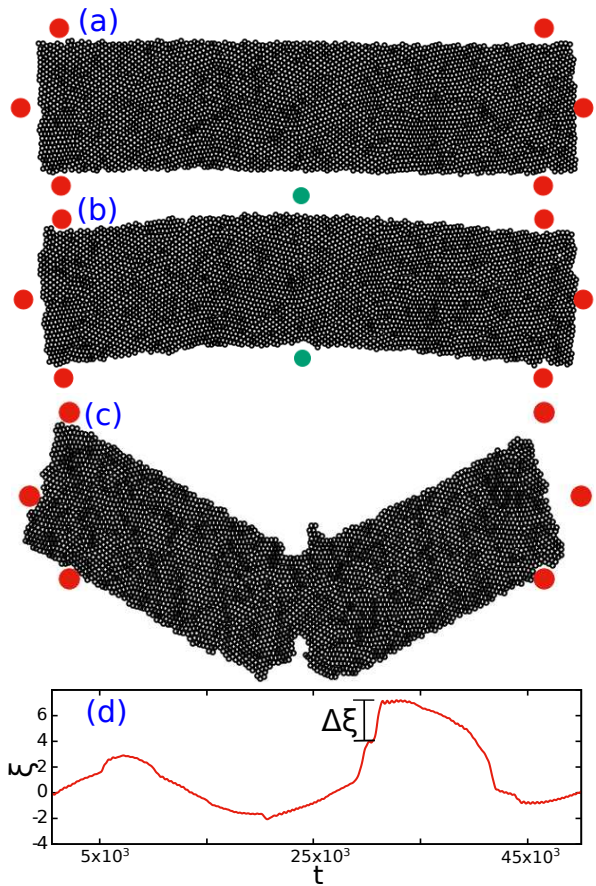}
\caption{Panel (a): the strip of system A with the stoppers and pusher. In red dots we denote the stoppers
that cannot be moved. The green dot is the pusher on which the force $F_{\rm app}$ is applied. Panel (b): the bent strip. Panel (c): The strip being broken after $n_f$ cycles. 
Panel (d): A typical trajectory of the center a mass of the strip.}
\label{diagram}
\end{figure}

We first equilibrate the ternary mixture which is in a rectangular box with periodic boundary conditions at high temperature $T = 1.0$, followed by NVT cooling down to $T = 0$ with a cooling rate $\dot{T} = 10^{-4}$. Once there, the system is equilibrated again at constant $N$, $P$ and $T$ by keeping $T = 0$ and $P = 0$. After that we remove the periodic boundaries. Finally we warm up to a very low temperature of $T=0.02$. We chose this small but finite temperature because at $T=0$ repeated bending tends to fall on limit cycles which we deem unphysical \cite{13RLR}. A typical simulation employs
a strip whose length (in the $x$ direction) and width (in the $y$ direction) are $L$ and $W$ respectively. The geometry and the placement of ``pushers" and "stoppers" are shown in Fig.~\ref{diagram}. This arrangement is motivated by the experiment of Bonn et al.~\cite{98BKPBM}.
In equilibrium the center of mass is at the origin. To bend the strip in the $\hat{y}$ and $-\hat{y}$ direction we apply a force $F$ at $(0, -W)$ and $(0, W)$ respectively. To prevent  a translational motion of the strip we use four stoppers which have positions pinned at $(-L/2 + 3\sigma_{st}, W/2 + \sigma_{st})$, $(L/2 - 3\sigma_{st}, W/2 + \sigma_{st})$, $(L/2 - 3\sigma_{st}, -W/2 - \sigma_{st})$, $(L/2 + 3\sigma_{st}, -W/2 - \sigma_{st})$, where $\sigma_{st}$ is the diameter of the stopper. We chose $\sigma_{st} = 5\sigma_{AA}$. A pusher particle with diameter $\sigma_{push} = 4\sigma_{AA}$ is used to apply the external force. To bend the strip in $\hat{y}$ direction we place the pusher at $(0, -W/2)$ and move it in $\hat{y}$ direction by increasing an external force $\B F(t) = F_{\rm app}\sin(\alpha t) \hat{y}$, where $F_{\rm app}$ is the maximum value of applied force. We choose $\alpha = (\frac{\pi}{2})\frac{10^{-3}}{F_{\rm app}}$, such that after a time $t = 10^{3}$ the force increases by one unit. The pusher attains the maximum value $F_{\rm app}$ after traversing a path of length $\xi_{\rm max}$. Then the external force is reduced back to 0 so the the strip can relax and return to its initial state. After that the pusher is moved to $(0, W/2)$ and we apply an external force in the $-\hat{y}$ direction. Now $F(t) = - F_{\rm app}\sin(\alpha t) \hat{y}$, and the pusher reaches its utmost negative position after traversing a length $\xi{\rm min}$. Finally the force is reversed again until the force vanishes: one cycle is then completed. These cycles are repeated $n$ times until the strip fails and breaks at some value $n=n_f$.
This bending protocol was applied to two different systems: system A has $N=4228$, $L=130$ and $W=24.4$. System B has $N = 3200$, $L=112$ and $W=21.0$. The bending protocol was applied to both systems at $T = 0.02$.  Note that the aspect ratio was kept constant at ($L/W \sim 5.33$). For system A we used seven different values of applied external force, $F_{app}$ = 6.25, 6.5, 6.75, 7.0, 7.5, 8.0 and 8.5. For system B we used
the seven values $F_{app}$ = 5.75, 6.0, 6.25, 6.5, 7.0, 7.25, 7.5.

{\bf Definition of measured quantities}:  Of primary interest in such fatigue experiments is the number of cycles $n_f$ at which the system fails. ``Failure" here is defined as the appearance of macroscopic break in the strip that is not healed by further bending cycles, see Fig.~\ref{diagram} panel (c). As expected, $n_f$ is a stochastic variable that is widely distributed, with very large sample to sample fluctuations even for one chosen value of $F_{\rm app}$. It therefore makes sense to focus on the distribution function $P(n_f;F_{\rm app})$ and its mean value, denoted as $\langle n_f \rangle(F_{\rm app}) $; both these quantities depend on $F_{\rm app}$. Two other quantities of interest are (i) the value of $F_{\rm app}$ that results in system's failure in {\em one} cycle. This value is denoted below as $F_y$ and it depends (for a given system size) on the temperature $T$; and (ii) the value of $F_{\rm app}$ below which we find no failure on the time scale of our simulations \cite{19Gon}. This value is denoted below as $F_L$. The W\"ohler plots presented below pertain to the range $F_L < F_{\rm app} < F_y$. 

The other quantity of major interest is the ``damage" $D_n^s$ that accumulates in every cycle. The definition of damage is not a-priori obvious. We propose here to define the damage as
the energy ``wasted" during plastic events. To introduce this quantity one observes the trajectory of the strip's center of mass, cf Fig.~\ref{diagram} panel (d). The trajectory is smooth modulo temperature fluctuations, but every now and then it suffers a discontinuity when a plastic event is taking place. The trajectory ``jumps" an amount $\Delta \xi$ within
and interval of time $\Delta t$ with 
the force $F(t)$ being fixed. To distinguish from temperature fluctuations we have employed a threshold of $\Delta \xi/\Delta t \ge 0.04$. These jumps were identified as the 
damage $F(t) \Delta \xi$ where the value of $F(t)$ was taken from the middle
of the interval $\Delta t$. The damage was added up for all the jumps occurring during
a given $n$th cycle, giving rise to the quantity $D^s_n$,
\begin{equation}
	D_n^s \equiv \sum_k F_k(t) (\Delta \xi)_k
\end{equation}
where the sum on $k$ runs on all the jumps taking place in the $n$th cycle. Finally we are interested in $D_{\rm acc}$ which is defined as the total accumulated damage
during all the cycles until collapse,
\begin{equation}
D_{\rm acc} (n_f) \equiv \sum_{n=1}^{n_f} 	D_n^s \ .
\label{defDacc}
\end{equation}  
This quantity will turn out to be crucial for the understanding of the collapse due to accumulated damage. We note here that small plastic events cannot be safely separated from temperature fluctuations, and therefore our measurements of the damage should be taken as a lower bound on the actual amount of energy spent on plastic events. 

\begin{figure}
	\includegraphics[scale=0.45]{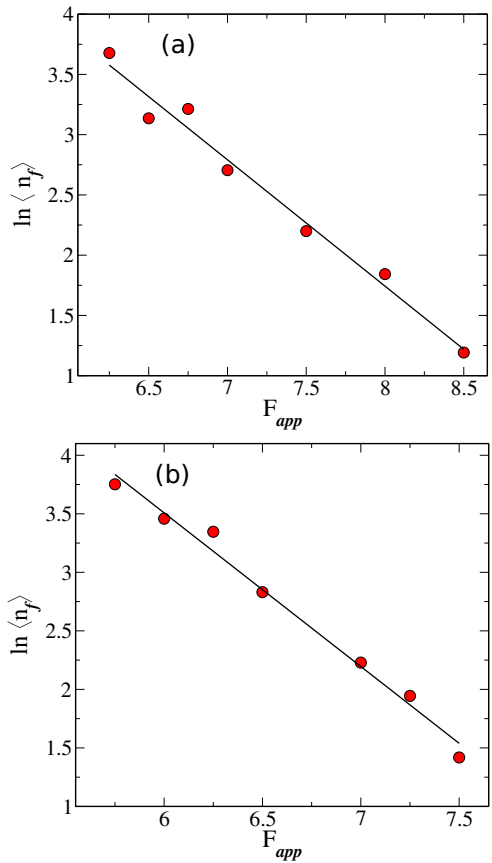}
	\caption{The W\"ohler plot for the amorphous strip. Panel (a): system A with $N=4228$.
	Panel (b): system B $N=3200$.}
	\label{wohler}
\end{figure}

{\bf Results of numerical simulations:} As mentioned above, the most typical measurement in engineering contexts is provided by the W\"ohler diagram, relating the number of cycles to failure to the stress level. 
In Fig.~\ref{wohler} we present the average number of cycles to failure $\langle n_f\rangle$ as a function of $F_{\rm app}$ in a log-linear plot for both systems A and B. The average was computed from about 500 realizations for every value of $F_{\rm app}$.
The data support an exponential dependence of the form
\begin{equation}
	\ln \langle n_f\rangle \approx a F_y -a F_{\rm app} \ ,
	\label{nf}
\end{equation}
where $F_{\rm y}$ is the applied force that breaks the system in one cycle, and the coefficient $a$ is dimensional, with units of inverse force, to be identified below. For systems A and B $aF_{\rm y}$ is $10.13\pm 0.48$  and $11.38\pm 0.45$ respectively. The numerical values of $a$ are  $a\approx 1.05\pm 0.07$ and $a\approx 1.31 \pm 0.07$ respectively. For future reference the parameter of this and further numerical fits are collected in Table~\ref{table}.

\begin{figure}
	\includegraphics[scale=0.60]{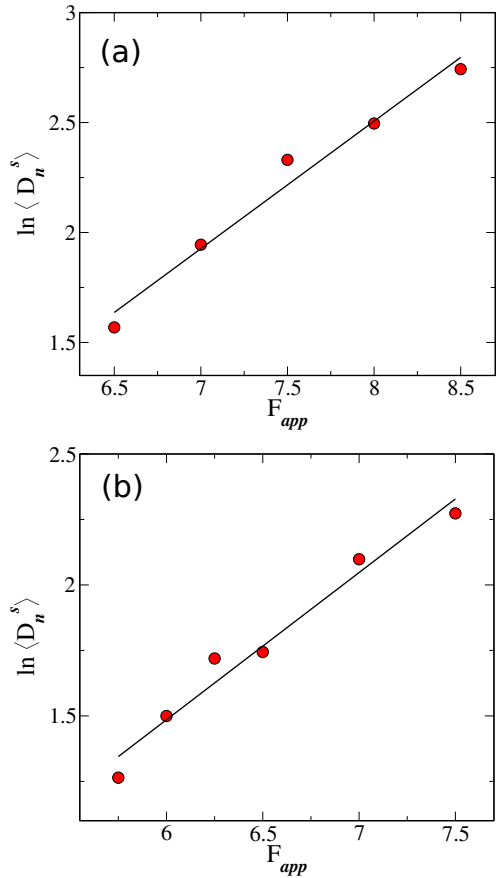}
	\caption{Dependence of the average damage per cycle on the maximal applied force. Panels (a) and (b) are for system A and B respectively.}
	\label{Ds}
\end{figure}

A first clue to the origin of Eq.~(\ref{nf}) is provided by the average of the damage per cycle
$\langle D^s_n\rangle$ and its dependence on $F_{\rm app}$, as shown in Fig.~\ref{Ds}.
This is also an exponential, {\em growing} with $ F_{\rm app}$;
\begin{equation}
	\ln \langle D^s_n\rangle \approx D_0+b F_{\rm app} \ ,
	\label{Dsexo}
\end{equation}
where $D_0=-2.13\pm 0.40$ and -1.89 $\pm 0.33$;  $b=0.58\pm 0.05$ and 0.56$\pm 0.05$ for systems A and B respectively. Here again $b$ is a constant with dimension of inverse force.
\begin{table}[h!]
	\begin{center}
		\begin{tabular}{c|c|c|c}
			~&	a & b & c \\
			\hline
			System A & $1.05\pm 0.07$ & $0.58\pm 0.05$ & $0.6\pm 0.04$  \\ \hline
			System B & $1.31\pm 0.07$ & $0.56\pm 0.05$ &$ 0.64\pm 0.03$  \\ \hline
		\end{tabular}
	\end{center}
	\caption{The parameters of dimension of inverse force that determine the exponential dependence on the applied force, see text for details.}
	\label{table}
\end{table}

\begin{figure}
	\vskip 0.4 cm
		\includegraphics[scale=0.50]{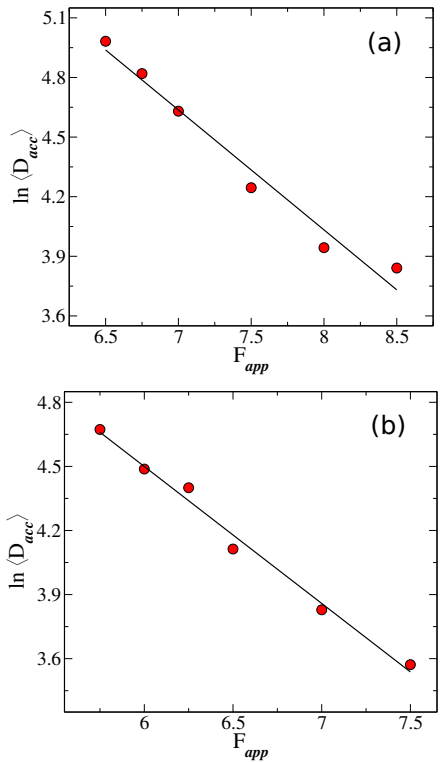}
	\caption{The dependence of $\langle D_{\rm acc} \rangle$ on $F_{\rm app}$. Systems A and B are in 
	panels (a) and (b) respectively.}
	\label{Dacc}
\end{figure}

Probably the most interesting numerical finding has to do with $D_{\rm acc}$. We find that with the exception of fragile configurations that break very quickly, for all the strips that survive more than about 20 cycles, $D_{\rm acc}$ attains a constant value that depends on $F_{\rm app}$ but not on the number of cycles \cite{07HDJS}. This result will have significant implications as it underlines the fact that accumulated damage is the critical physical quantity that leads to mechanical collapse. The dependence of the average of this quantity on $F_{\rm app}$ is shown in Fig.~\ref{Dacc}. Computing the average $\langle D_{\rm acc} \rangle$ the data support again an exponential fit of the form
\begin{equation}
\ln \langle D_{\rm acc} \rangle \approx C  -c F_{\rm app}  \ , 
\label{Daccexp}
\end{equation}
Here $C = 8.86\pm 0.38$ and 8.34$\pm 0.23$ and  $c=0.6\pm 0.04$ and 0.64$\pm 0.03$ respectively for systems A and B. The constant $c$ is the last  parameter with dimension of inverse force. 

{\bf Scaling theory:} In the rest of this Letter we offer a scaling theory to rationalize the numerical values of the coefficients $a$, $b$ and $c$ which are collected in Table~\ref{table}. This theory is tentative, and further simulations and experimental tests are called for its establishment. We propose it here as a first step in the solution of the long standing riddle of W\"ohler diagrams.

The first observation is that Eqs.~(\ref{Dsexo}) and (\ref{Daccexp}) provide
a physical reason for the W\"ohler relation Eq.~(\ref{nf}). The idea is that the average number
of cycles to failure will be determined by the following ratio:
\begin{equation}
	\langle n_f \rangle \sim  \langle D_{\rm acc}\rangle /	\langle D_n^s\rangle \ ,
	\label{explain}	
\end{equation}
up to a constant of the order of unity. Simply the amount of damage per cycle accumulates to the (approximately) constant
value $\langle D_{\rm acc}\rangle$ during $\langle n_f\rangle $ cycles. We realize that scaling theory can only provide predictions up to constants of the order of unity, but nevertheless it is worthwhile to see how well we can explain the numerical results. Plugging in Eq.~(\ref{explain}) the
numerical values of the pre-exponential constants and the values of $a$ and $b$ for both systems we estimate
\begin{eqnarray}
\ln	\langle n_f \rangle &\approx& 10.99\pm 0.78 - (1.18\pm 0.09) F_{\rm app}\ , \text{system A} \
\\
\ln	\langle n_f \rangle &\approx& 10.23\pm 0.56 - (1.2\pm 0.08) F_{\rm app}\ , \text{system B}. \nonumber
\label{comparison}
\end{eqnarray} 
These result are consistent within the error bars with the numerical  simulations Eq.~(\ref{nf}).  We therefore can propose a scaling relation: 
\begin{equation}
	a=b+c\ .
	\label{wow}
\end{equation}

We reiterate at this point that all these three numbers ($a, b$ and $c$) are dimensional, being inverse forces. Until now we do not have any prior knowledge of these dimensional coefficients. To seek this information we examine again the values of $\langle D_{\rm acc} \rangle$ and realize that they span one order of $e$. We will therefore construct an approximate scale
of damage by averaging $\langle D_{\rm acc} \rangle$ over its range $F_L\le F_{\rm app} \le F_y$, creating an average of averages, denoted as 
$\overline{\langle {D_{\rm acc}} \rangle}$. From the numerics we measure $\overline{\langle {D_{\rm acc}} \rangle}\approx 111$ for system A and
88 for system B.

With this scale in mind, consider the parameter $b$ in Eq.~(\ref{Dsexo}). It has the dimension of inverse force and must be independent of $F_{\rm app}$. The only energy scale available that is independent of $F_{\rm app}$ is  $\overline{\langle {D_{\rm acc}} \rangle}$, and the length scale associated with damage must be $\ell_D\equiv \sqrt{LW}$, since plastic events can appear anywhere in the area of the strip. For system A
$\ell_D=56.36$ whereas for system B $\ell_D=48.5$ . We thus estimate
\begin{eqnarray}
b\approx \ell_D/\overline{\langle D_{\rm acc} \rangle} &\approx& 0.51 \quad \text{system A} \ ,\\
&\approx& 0.55 \quad \text{system B} \ .
\label{amazing}	
\end{eqnarray}
Taking into account the approximate nature of the scale $\overline{\langle {D_{\rm acc}} \rangle}$ and the fact that it is a lower bound, we consider the results to be in good agreement with the data. 

Regarding the numerical value of $a$, we expect that failure starts with a micro-crack at the upper or lower boundary, so the relevant scale for Eq.~(\ref{nf}) is $L$. Therefore we estimate
\begin{eqnarray}
	a\approx L/\overline{\langle D_{\rm acc} \rangle} &\approx& 1.17 \quad \text{system A} \ ,\\
&\approx& 1.27 \quad \text{system B} \ .
	\label{nice}	
\end{eqnarray}
which is again in fair agreement with the data.

In summary, we offer a scaling theory of W\"ohler plots, based on the idea that accumulated damage is the fundamental cause for failure, joined with the discovery that this quantity appears constant for system failing at different values of $\langle n_f \rangle$. This phenomenological finding indicates that memory plays a crucial role in fatigue; one can fail after many cycle if the damage accrued in each cycle is small, or in a few cycles if the damage in each happens to be large. What matters is the accumulated damage that seems to have a (presumably material and temperature dependent) limit \cite{07HDJS}. The measurement of this quantity is not trivial due to the difficulty of distinguishing small plastic events from temperature fluctuations. Nevertheless, it appears that focusing on this quantity allows understanding of the exponential dependence of the average number of cycles to failure on the applied force. Moreover, introducing the natural scales for failure on the boundary and plastic events in the bulk, the numerical values of the parameters $a$, $b$ and $c$ could be rationalized. 

It would be very useful to test the ideas presented in this Letter in experiments. The experimenters will need to come up with a robust method to estimate the damage done in each cycle and its accumulated counterpart. If it turned out that the accumulated damage is indeed  independent of the average number of cycles for failure, the path for a scaling theory of the type presented here would open. In parallel, in future work the present effort would continue using numerical simulations and additional theoretical developments.  

\acknowledgments This work has been supported in part by the Minerva Foundation, Munich, Germany, through the Minerva Center for Aging at the Weizmann Institute of Science. We thank Jacques Zylberg for his involvement in the initiation of this project.

\bibliography{bent.strip}

\begin{thebibliography}{11}%
\makeatletter
\providecommand \@ifxundefined [1]{%
 \@ifx{#1\undefined}
}%
\providecommand \@ifnum [1]{%
 \ifnum #1\expandafter \@firstoftwo
 \else \expandafter \@secondoftwo
 \fi
}%
\providecommand \@ifx [1]{%
 \ifx #1\expandafter \@firstoftwo
 \else \expandafter \@secondoftwo
 \fi
}%
\providecommand \natexlab [1]{#1}%
\providecommand \enquote  [1]{``#1''}%
\providecommand \bibnamefont  [1]{#1}%
\providecommand \bibfnamefont [1]{#1}%
\providecommand \citenamefont [1]{#1}%
\providecommand \href@noop [0]{\@secondoftwo}%
\providecommand \href [0]{\begingroup \@sanitize@url \@href}%
\providecommand \@href[1]{\@@startlink{#1}\@@href}%
\providecommand \@@href[1]{\endgroup#1\@@endlink}%
\providecommand \@sanitize@url [0]{\catcode `\\12\catcode `\$12\catcode
  `\&12\catcode `\#12\catcode `\^12\catcode `\_12\catcode `\%12\relax}%
\providecommand \@@startlink[1]{}%
\providecommand \@@endlink[0]{}%
\providecommand \url  [0]{\begingroup\@sanitize@url \@url }%
\providecommand \@url [1]{\endgroup\@href {#1}{\urlprefix }}%
\providecommand \urlprefix  [0]{URL }%
\providecommand \Eprint [0]{\href }%
\providecommand \doibase [0]{http://dx.doi.org/}%
\providecommand \selectlanguage [0]{\@gobble}%
\providecommand \bibinfo  [0]{\@secondoftwo}%
\providecommand \bibfield  [0]{\@secondoftwo}%
\providecommand \translation [1]{[#1]}%
\providecommand \BibitemOpen [0]{}%
\providecommand \bibitemStop [0]{}%
\providecommand \bibitemNoStop [0]{.\EOS\space}%
\providecommand \EOS [0]{\spacefactor3000\relax}%
\providecommand \BibitemShut  [1]{\csname bibitem#1\endcsname}%
\let\auto@bib@innerbib\@empty
\bibitem [{\citenamefont {Miles}(1954)}]{54Mil}%
  \BibitemOpen
  \bibfield  {author} {\bibinfo {author} {\bibfnamefont {J.~W.}\ \bibnamefont
  {Miles}},\ }\href@noop {} {\bibfield  {journal} {\bibinfo  {journal} {Journal
  of the Aeronautical Sciences}\ }\textbf {\bibinfo {volume} {21}},\ \bibinfo
  {pages} {753} (\bibinfo {year} {1954})}\BibitemShut {NoStop}%
\bibitem [{\citenamefont {Dowell}(1966)}]{66Dow}%
  \BibitemOpen
  \bibfield  {author} {\bibinfo {author} {\bibfnamefont {E.~H.}\ \bibnamefont
  {Dowell}},\ }\href {\doibase 10.2514/3.3658} {\bibfield  {journal} {\bibinfo
  {journal} {AIAA Journal}\ }\textbf {\bibinfo {volume} {4}},\ \bibinfo {pages}
  {1267} (\bibinfo {year} {1966})}\BibitemShut {NoStop}%
\bibitem [{\citenamefont {Dietmann}\ \emph {et~al.}(1989)\citenamefont
  {Dietmann}, \citenamefont {Bhongbhibhat},\ and\ \citenamefont
  {Schmid}}]{89DBS}%
  \BibitemOpen
  \bibfield  {author} {\bibinfo {author} {\bibfnamefont {H.}~\bibnamefont
  {Dietmann}}, \bibinfo {author} {\bibfnamefont {T.}~\bibnamefont
  {Bhongbhibhat}}, \ and\ \bibinfo {author} {\bibfnamefont {A.}~\bibnamefont
  {Schmid}},\ }in\ \href@noop {} {\emph {\bibinfo {booktitle} {ICBMFF3}}}\
  (\bibinfo {year} {1989})\BibitemShut {NoStop}%
\bibitem [{\citenamefont {Klevtsov}\ and\ \citenamefont {Crane}(1994)}]{94KC}%
  \BibitemOpen
  \bibfield  {author} {\bibinfo {author} {\bibfnamefont {I.}~\bibnamefont
  {Klevtsov}}\ and\ \bibinfo {author} {\bibfnamefont {R.}~\bibnamefont
  {Crane}},\ }\href {\doibase 10.1115/1.2929563} {\bibfield  {journal}
  {\bibinfo  {journal} {Journal of Pressure Vessel Technology}\ }\textbf
  {\bibinfo {volume} {116}},\ \bibinfo {pages} {110} (\bibinfo {year}
  {1994})}\BibitemShut {NoStop}%
\bibitem [{\citenamefont {Freudenthal}\ \emph {et~al.}(1953)\citenamefont
  {Freudenthal}, \citenamefont {Gumbel},\ and\ \citenamefont {Gough}}]{53FGG}%
  \BibitemOpen
  \bibfield  {author} {\bibinfo {author} {\bibfnamefont {A.~M.}\ \bibnamefont
  {Freudenthal}}, \bibinfo {author} {\bibfnamefont {E.~J.}\ \bibnamefont
  {Gumbel}}, \ and\ \bibinfo {author} {\bibfnamefont {H.~J.}\ \bibnamefont
  {Gough}},\ }\href {\doibase 10.1098/rspa.1953.0024} {\bibfield  {journal}
  {\bibinfo  {journal} {Proc. R. Soc. A Math. Phys. Eng. Sci.}\ }\textbf
  {\bibinfo {volume} {216}},\ \bibinfo {pages} {309} (\bibinfo {year}
  {1953})}\BibitemShut {NoStop}%
\bibitem [{\citenamefont {Parmar}\ \emph {et~al.}(2020)\citenamefont {Parmar},
  \citenamefont {Ozawa},\ and\ \citenamefont {Berthier}}]{20POB}%
  \BibitemOpen
  \bibfield  {author} {\bibinfo {author} {\bibfnamefont {A.~D.}\ \bibnamefont
  {Parmar}}, \bibinfo {author} {\bibfnamefont {M.}~\bibnamefont {Ozawa}}, \
  and\ \bibinfo {author} {\bibfnamefont {L.}~\bibnamefont {Berthier}},\
  }\href@noop {} {\bibfield  {journal} {\bibinfo  {journal} {Physical Review
  Letters}\ }\textbf {\bibinfo {volume} {125}},\ \bibinfo {pages} {085505}
  (\bibinfo {year} {2020})}\BibitemShut {NoStop}%
\bibitem [{\citenamefont {Dauchot}\ \emph {et~al.}(2011)\citenamefont
  {Dauchot}, \citenamefont {Karmakar}, \citenamefont {Procaccia},\ and\
  \citenamefont {Zylberg}}]{11DKPZ}%
  \BibitemOpen
  \bibfield  {author} {\bibinfo {author} {\bibfnamefont {O.}~\bibnamefont
  {Dauchot}}, \bibinfo {author} {\bibfnamefont {S.}~\bibnamefont {Karmakar}},
  \bibinfo {author} {\bibfnamefont {I.}~\bibnamefont {Procaccia}}, \ and\
  \bibinfo {author} {\bibfnamefont {J.}~\bibnamefont {Zylberg}},\ }\href
  {\doibase 10.1103/PhysRevE.84.046105} {\bibfield  {journal} {\bibinfo
  {journal} {Phys. Rev. E}\ }\textbf {\bibinfo {volume} {84}},\ \bibinfo
  {pages} {046105} (\bibinfo {year} {2011})}\BibitemShut {NoStop}%
\bibitem [{\citenamefont {Regev}\ \emph {et~al.}(2013)\citenamefont {Regev},
  \citenamefont {Lookman},\ and\ \citenamefont {Reichhardt}}]{13RLR}%
  \BibitemOpen
  \bibfield  {author} {\bibinfo {author} {\bibfnamefont {I.}~\bibnamefont
  {Regev}}, \bibinfo {author} {\bibfnamefont {T.}~\bibnamefont {Lookman}}, \
  and\ \bibinfo {author} {\bibfnamefont {C.}~\bibnamefont {Reichhardt}},\
  }\href {\doibase 10.1103/PhysRevE.88.062401} {\bibfield  {journal} {\bibinfo
  {journal} {Phys. Rev. E}\ }\textbf {\bibinfo {volume} {88}},\ \bibinfo
  {pages} {062401} (\bibinfo {year} {2013})}\BibitemShut {NoStop}%
\bibitem [{\citenamefont {Bonn}\ \emph {et~al.}(1998)\citenamefont {Bonn},
  \citenamefont {Kellay}, \citenamefont {Prochnow}, \citenamefont
  {Ben-Djemiaa},\ and\ \citenamefont {Meunier}}]{98BKPBM}%
  \BibitemOpen
  \bibfield  {author} {\bibinfo {author} {\bibfnamefont {D.}~\bibnamefont
  {Bonn}}, \bibinfo {author} {\bibfnamefont {H.}~\bibnamefont {Kellay}},
  \bibinfo {author} {\bibfnamefont {M.}~\bibnamefont {Prochnow}}, \bibinfo
  {author} {\bibfnamefont {K.}~\bibnamefont {Ben-Djemiaa}}, \ and\ \bibinfo
  {author} {\bibfnamefont {J.}~\bibnamefont {Meunier}},\ }\href@noop {}
  {\bibfield  {journal} {\bibinfo  {journal} {Science}\ }\textbf {\bibinfo
  {volume} {280}},\ \bibinfo {pages} {265} (\bibinfo {year}
  {1998})}\BibitemShut {NoStop}%
\bibitem [{\citenamefont {Gonz{\'a}lez-Vel{\'a}zquez}(2019)}]{19Gon}%
  \BibitemOpen
  \bibfield  {author} {\bibinfo {author} {\bibfnamefont {J.~L.}\ \bibnamefont
  {Gonz{\'a}lez-Vel{\'a}zquez}},\ }\href@noop {} {\emph {\bibinfo {title}
  {Mechanical behavior and fracture of engineering materials}}}\ (\bibinfo
  {publisher} {Springer},\ \bibinfo {year} {2019})\BibitemShut {NoStop}%
\bibitem [{\citenamefont {Harmon}\ \emph {et~al.}(2007)\citenamefont {Harmon},
  \citenamefont {Demetriou}, \citenamefont {Johnson},\ and\ \citenamefont
  {Samwer}}]{07HDJS}%
  \BibitemOpen
  \bibfield  {author} {\bibinfo {author} {\bibfnamefont {J.~S.}\ \bibnamefont
  {Harmon}}, \bibinfo {author} {\bibfnamefont {M.~D.}\ \bibnamefont
  {Demetriou}}, \bibinfo {author} {\bibfnamefont {W.~L.}\ \bibnamefont
  {Johnson}}, \ and\ \bibinfo {author} {\bibfnamefont {K.}~\bibnamefont
  {Samwer}},\ }\href@noop {} {\bibfield  {journal} {\bibinfo  {journal}
  {Physical Review Letters}\ }\textbf {\bibinfo {volume} {99}},\ \bibinfo
  {pages} {135502} (\bibinfo {year} {2007})}\BibitemShut {NoStop}%
\end{thebibliography}%

\end{document}